\documentclass[11pt]{aastex}
\usepackage{emulateapj5,apjfonts}
\usepackage{epsf}

\def\teff{T$_{\rm{eff}}$}

\def\etal{{et~al.\,}}

\def\mj{$M_{\rm J}\,$}
\def\rj{$R_{\rm J}\,$}

\def\Dwa{$\,$\uppercase\expandafter{\romannumeral5}$\,$}
\def\mic{$\mu$m$\,$}

\def\sless{\lower2pt\hbox{$\buildrel {\scriptstyle <}
   \over {\scriptstyle\sim}$}}
\def\sgreat{\lower2pt\hbox{$\buildrel {\scriptstyle >}
   \over {\scriptstyle\sim}$}}
\def\aa{Astron. Astrophys.\ }
\def\apjl{Astrophys.~J.~Letters\ }

\begin{document}

\title{Spectra and Diagnostics for the Direct Detection of Wide-Separation Extrasolar Giant Planets}

\author{Adam Burrows\altaffilmark{1}, David Sudarsky\altaffilmark{1}, \&  Ivan Hubeny\altaffilmark{1,2,3}}

\altaffiltext{1}{Department of Astronomy and Steward Observatory, 
                 The University of Arizona, Tucson, AZ \ 85721}
\altaffiltext{2}{NOAO, Tucson, AZ 85726}
\altaffiltext{3}{NASA Goddard Space Flight Center, Greenbelt, MD 20771}
\begin{abstract}

We calculate as a function of orbital distance, mass, 
and age the theoretical spectra and orbit-averaged planet/star flux ratios for
representative wide-separation extrasolar giant planets (EGPs)
in the optical, near-infrared, and mid-infrared.  
Stellar irradiation of the planet's atmosphere
and the effects of water and ammonia clouds
are incorporated and handled in a consistent fashion.
We include predictions for 12 specific known EGPs.
In the process, we derive physical diagnostics that can inform the
direct EGP detection and remote sensing 
programs now being planned or proposed.  Furthermore, we calculate
the effects of irradiation on the spectra of a representative companion 
brown dwarf as a function of orbital distance.

\end{abstract}

\section{Introduction}
\label{intro}

To date, more than 110 EGPs (Extrasolar Giant Planets) have been discovered by the radial-velocity
technique around stars with spectral types from M4 to F7\footnote{see J. Schneider's 
Extrasolar Planet Encyclopaedia at http://www.obspm.fr/encycl/encycl.html
for a reasonable listing, with comments.}.  A sample of 
relevant references to the discovery literature, by no means exhaustive,
includes Mayor and Queloz (1995), Marcy and Butler (1996), Butler et al.
(1997,1999), Marcy et al. (1998,1999), Marcy, Cochran, and Mayor (2000),
Queloz et al. (2000), Santos et al. (2000), and Konacki \etal (2003).
These planets have minimum masses ($m_p\sin(i)$, where $i$ is the orbital inclination) between $\sim$0.12 \mj
and $\sim$15 \mj (\mj $\equiv$ one Jupiter mass), orbital semi-major axes 
from $\sim$0.0225 AU to $\sim$5.9 AU, and eccentricities from $\sim$0 to above 0.7.
Given our all-too-narrow experience within the solar system, 
such variety and breadth was wholly unanticipated.

Importantly, two EGPs (HD 209458b and OGLE-TR-56b)
have been found to transit their primaries (Henry
\etal 2000; Charbonneau \etal 2000,2001; Brown
\etal 2001; Konacki \etal 2003; Torres et al. 2003).
Furthermore, in the first measurement of the composition of an extrasolar
planet of any kind, Charbonneau et al. (2002) detected sodium (Na-D)
in their HD 209458b transit spectrum.  This was
followed by the detection of atomic hydrogen at Lyman-$\alpha$
and the discovery of a planetary wind (Vidal-Madjar et al. 2003; Burrows and Lunine 1995).
With both transit and radial-velocity data, an EGP's mass and radius
can be determined, enabling its physical and structural study.
Such data can resolve the ambiguity inherent in the radial-velocity technique's
sensitivity to only the combination $m_p\sin(i)$.  Precision astrometry can also be used
to derive masses, as has been done for GJ 876b with the Fine Guidance Sensors on HST 
(Bennett et al. 2002), and space interferometry using SIM (Unwin and Shao 2000) promises
to provide unprecedented astrometric masses by the year $\sim$2010.  

However, it is only by direct detection of a planet's light using photometry, spectrophotometry,
or spectroscopy that the detailed and rigorous study of its physical attributes can be conducted.
By this means, the composition, gravity, radius, and mass of the giant might be derived and the general theory
of EGP properties and evolution might be tested (Burrows et al. 1995,1997; Marley et al. 1999; Sudarsky, 
Burrows and Pinto 2000; Sudarsky, Burrows, and Hubeny 2003 (SBH); Baraffe et al. 2003).
For close-in EGPs, we can anticipate in the next few years wide-band precision photometry  
from MOST (Matthews et al. 2001), Kepler (Koch et al. 1998), Corot (Antonello and Ruiz 2002), or MONS (Christensen-Dalsgaard 2000)
that will provide the variations of the summed light of the planet and star due to changes in the planetary phase.
In the mid- to far-infrared, the Spitzer Space Telescope (a.k.a. SIRTF, Space InfraRed Telescope Facility; Werner and Fanson 1995)
might soon be able to measure the variations in the planet/star flux ratios of close-in EGPs (SBH). 
For wide-separation EGPs, it is necessary to measure the planet's light from under the 
glare of the primary star at very high star-to-planet contrast ratios.  To achieve 
this from the ground, telescopes such as the VLT interferometer (Paresce 2001),
the Keck interferometer (van Belle and Vasisht 1998; Akeson and Swain 2000; Akeson, 
Swain, and Colavita 2001), and the LBT nulling interferometer (Hinz 2001)
will be enlisted.  From space, the Terrestrial Planet Finder (TPF, 
Levine et al. 2003) and/or a coronagraphic optical imager such as {\it Eclipse} (Trauger et al. 2000,2001) 
could obtain low-resolution spectra. 

To support the above efforts and planning for future programs of direct 
detection of extraoslar giant planets and to provide the theoretical context for
the general analysis of the spectra and photometry of irradiated and isolated EGPs, our group has embarked upon 
a series of papers of EGP spectra, evolution, chemistry, transits, orbital phase functions, and light curves.
The most recent paper in this series (SBH) explored generic features
of irradiated EGP spectra as a function of orbital distance, cloud properties, and composition class (Sudarsky, Burrows, and Pinto 2000).   
Our technical approach vis \`a vis radiative transfer, molecular abundance determinations, and cloud
modeling is described in detail in that paper, to which the interested reader is referred.  
However, SBH did not explore the diagnostics of planetary mass and age.  In addition, in their
study of the orbital distance dependence of EGP spectra SBH did not incorporate the effects of water or ammonia clouds 
in a fully consistent fashion.  

In the current paper, we allow our atmosphere code to determine 
cloud placement in a fully iterative, converged fashion. 
The result is a consistent determination of the dependence 
on distance of the spectra of EGPs irradiated by a G2V star,
including the effects of the water and ammonia clouds that should form in their atmospheres. 
In \S\ref{numerical}, we summarize our numerical approach.  Then in \S\ref{distance_sequence}
we present and describe our results for the dependence of the spectra of irradiated EGPs on orbital distance.
In this paper, we emphasize the results for EGPs at wider separations ($>0.2$ AU) and defer discussion
of the corresponding theory for close-in EGPs to a later paper\footnote{Note, however, that the spectra of close-in EGPs 
has been addressed in SBH, as well as in Seager and Sasselov (1998,2000), Seager, Whitney, and Sasselov (2000), and Goukenleuque et al. 2000.}.
This section treats the entire expected range of EGP emission/reflection spectra and behavior.
In \S\ref{age_sequence}, we provide a representative sequence of models that portray the dependence of irradiated EGP spectra 
on age, at a given mass and orbital distance.  In \S\ref{mass_sequence}, we present a representative sequence with
EGP mass, at a given age and orbital distance.  Section \ref{browndwarfs} is a digression into the effect of stellar irradiation
on companion brown dwarfs, characterized by much larger masses and more slowly decaying heat content.  We study
the signature in the optical of the reflection of stellar light from a companion brown dwarf.
In fact, much of this paper is concerned with the signatures and diagnostics of the physical parameters
of irradiated substellar-mass companions\footnote{We refer to Substellar-Mass Objects as SMOs.}.  However, it is not feasible in one paper to explore all the 
possible combinations of planetary mass, age, composition, orbital semi-major axis, eccentricity, and orbital phase
with all stellar types.  Hence, to maintain a reasonable focus, we restrict our discussions to G2V primaries,
zero eccentricity orbits, and solar metallicity.  We narrow our scope further by plotting
only phase-averaged spectra (as in SBH) at zero orbital inclination.   
Papers on EGP orbital phase functions, 
albedos, and light curves are to follow this one (e.g., Sudarsky, Burrows, and 
Hubeny 2004). These papers address the dependence on phase and Keplerian parameters. 
Finally, in \S\ref{specific_EGPs} we present predicted phase-averaged 
planet/star flux ratios for several known EGPs
at wide separations and \S\ref{conclusion} reprises the 
essential conclusions of the paper.

\section{Numerical Issues} 
\label{numerical}

To calculate radiative/convective equilibrium atmospheres
and spectra we use a specific variant of the computer program TLUSTY (Hubeny 1988;
Hubeny \& Lanz 1995).  This variant involves the hybrid Complete
Linearization/Accelerated Lambda Iteration (CL/ALI)
method, although in the present runs we use a full ALI mode which
leads to an essential saving of computer time without
slowing down the iteration process significantly.
In addition, we employ TLUSTY's Discontinuous Finite Element (DFE)
version  (Castor, Dykema, \& Klein 1992).  The DFE technique, being 
first order, is optimal for handling irradiated atmospheres (SBH). 
All models are converged to one part in $10^3$.
Stellar spectra from Kurucz (1994)
are used for the incoming fluxes at the outer boundaries.  The inner boundary condition
is the interior flux and this (indexed by \teff) is taken from the
evolutionary models of Burrows et al. (1997) for the given mass and age, unless otherwise
indicated.  This approximate procedure works well for wider-separation EGPs,
but not as well for the closer-in EGPs ($<0.15$ AU).  We use a standard
mixing-length prescription to handle convection and a mixing length of
one pressure scale height.  Since the atmosphere code is planar, we use the redistribution
technique descibed in SBH, i.e. we weight the incident flux by 1/2 to
account for the average inclination of the planetary surface to the line of sight 
to the primary.  As described in SBH, the planetary spectra we present here are phase/time-averaged over the orbit. 
Care is taken with this procedure to ensure
that energy is conserved and that energy $in$ (from the star
and the planetary interior) equals energy $out$.  To account for the anisotropy of single scattering off of
cloud particles, we calculate using Mie theory the average of the cosine of the scattering angle,
and reduce the Mie-theory-derived total scattering cross section by one minus this average.
With this procedure, we are substituting the ``transport cross section" for the total
scattering cross section.  This approach has been shown to mimic the effect of asymmetric 
scattering quite well (Sudarsky, Burrows, and Pinto 2000).

For molecular and atomic compositions, we use an updated version of 
the chemical code of Burrows and Sharp (1999),
which includes a prescription to account for the rainout of condensed species in
a gravitational field and new thermochemical data.   The derived molecular abundances are very similar to those
obtained by Lodders (1999) and Lodders and Fegley (2002).  Our solar 
metallicity is defined as the elemental abundance pattern 
found in Anders and Grevesse (1989).  For molecular and atomic opacities, 
we have developed an extensive database, described in part in Burrows et al. (2001) and SBH.

The treatment of H$_2$O and NH$_3$ clouds is done in a manner consistent with their
respective condensation curves and the cloud base is put at the intersection
of the corresponding condensation curve at solar metallicity 
with the object's temperature/pressure ($T/P$) profile.
In each iteration of the global ALI scheme, we find
a position of the cloud base as the intersection of the
corresponding condensation curve with the current
$T/P$ profile. The scale height
of a cloud is assumed to be equal to one pressure scale
height. We use Mie theory for the absorptive and
scattering opacities of particles whose model size is
determined by the theory of Cooper et al. (2003).
In the subsequent iteration of the ALI scheme, we employ
this cloud opacity and scattering self-consistently in the
radiative transfer equation and the energy balance
equation. In the next ALI iteration we again recalculate
the position of the cloud base, and the whole process
is repeated until the cloud position is fully stabilized.
We note that in the initial stages of the
global iteration process the cloud position may vary
significantly; in some case clouds appear (disappear)
after several iterations of the cloudless (cloudy) atmosphere.
This procedure ensures that the cloud position is
self-consistent with the overall model atmosphere. This
also means that the predicted cloud properties, in particular 
optical depths and base pressures, vary
in a physical and consistent way with the orbital
distance, mass, and age.

\section{Orbital Distance Dependence of EGP Spectra from 0.2 to 15 AU}
\label{distance_sequence}

\begin{figure} 
\plotone{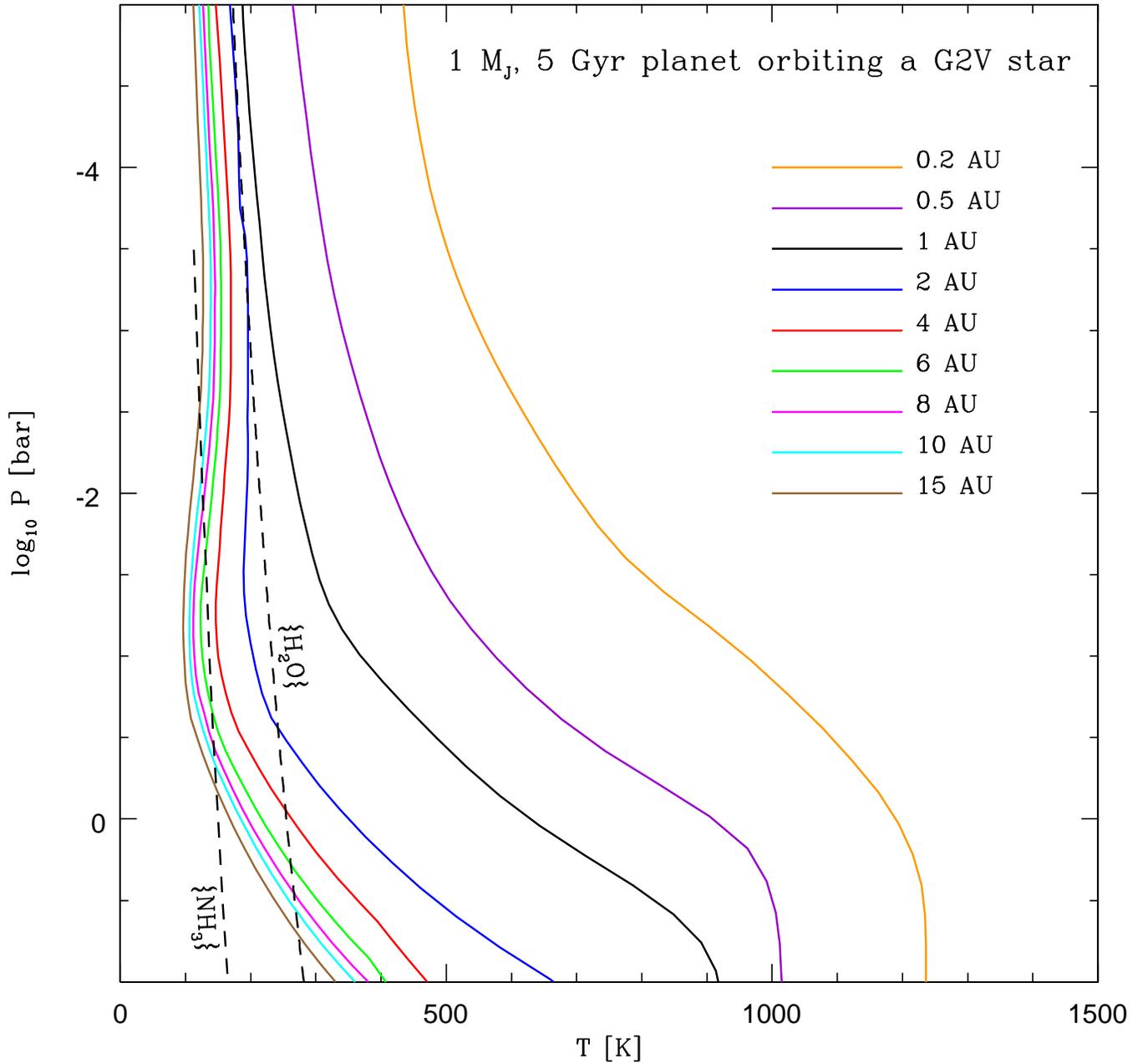}
\caption{Profiles of atmospheric temperature (in Kelvin) versus the logarithm base ten of the pressure (in bars)
for a family of irradiated 1-\mj EGPs around a G2V star as a function of orbital distance.
Note that the pressure is decreasing along the ordinate, which thereby resembles altitude.
The orbits are assumed to be circular, the planets are assumed to 
have a radius of 1 \rj, the effective temperature of the inner boundary flux is 
set equal to 100 K,  and the orbital separations vary from 0.2 AU to 15 AU.
The intercepts with the dashed lines identified with either \{NH$_3$\} or \{H$_2$O\} denote the  
positions where the corresponding clouds form. See text for a discussion.} 
\label{auseq.c}
\end{figure}

Figure \ref{auseq.c} shows the $T/P$ profiles for the distance sequence
from 0.2 to 15 AU for a 1-\mj EGP irradiated by a G2V star.  For specificity,
we have assumed a radius of 1 \rj and an internal flux \teff\ of 100 K for the entire family.  According to the models
of Burrows et al. (1997), this corresponds roughly to an age of 5 Gyr, but after $\sim$0.1 Gyr,
the radius and gravity of the EGP vary little.   
The orbits are taken to be circular, so there is no assumed orbital phase dependence
\footnote{Note that with a significant eccentricity, this assumption is not valid.}.
The intercepts with the dashed lines identified by either \{NH$_3$\} or \{H$_2$O\} denote the
positions where the corresponding clouds form. Due to the cold trap effect and depletion due to
rainout (Burrows and Sharp 1999), the higher-pressure intercept is taken to be at the base of the cloud. The
spectral/atmospheric models include the effects of these clouds in a consistent way.
Table \ref{data.dist} gives the modal particle sizes in microns that
we derive using the theory of Cooper et al. (2003) when a cloud of either water or ammonia (or both) appears.
For the ammonia clouds that form in this orbital distance sequence (at $\sgreat 6$ AU),
the modal particle sizes we find hover near 50-60 \mic.
The corresponding particle sizes in the water clouds are near 110 \mic.  These particles are larger
than for more massive SMOs.  Furthermore, we assume that 
the particle size does not vary with altitude and that the particle size distribution 
(given a modal radius) is that of Deirmendjian (1964,1969).  Clearly,  
a major ambiguity in EGP modeling is cloud physics (SBH).  We have settled on
the Cooper et al. (2003) theory to provide a consistent framework. 

When the chemistry indicates that both cloud types are present, 
we include them both in the atmospheric/spectral calculation.
As Fig. \ref{auseq.c} shows, the ammonia cloud is always above the water cloud.
Water clouds form around a G2V star exterior to a distance near 1.5 AU, whereas ammonia
clouds form around such a star exterior to a distance near 4.5 AU.  Note that Jupiter itself
is at the distance from the Sun of $\sim$5.2 AU.  For the closer-in objects, the $T/P$ profile
manifests an inflection.  This inflection is a consequence of the dominance of external radiation
over the internal heat flux.  For longer ages and low masses, the orbital distance at which one must
place the SMO/EGP to erase this inflection is large.  For this model set, that distance is $\sim$2 AU.
For larger masses, dimmer primaries, and shorter ages, that distance decreases.  In addition,
the strong irradiation that produces the inflection in the $T/P$ profile also forces
the radiative/convective boundary to recede to higher pressures.  For a 1-\mj EGP at an orbital
distance of 0.05 AU (not shown) around a G2V  star, this pressure can be greater than 1000 bars!
Such is the case for HD 209458b and OGLE-TR56b (Burrows, Sudarsky, and Hubbard 2003; Fortney et al. 2003).

For comparison, and in anticipation of the discussion in \S\ref{age_sequence} concerning
the age dependence of irradiated EGP spectra, Fig. \ref{ageseq.c}
portrays the evolution in the $T/P$ profile of a 1-\mj EGP at a distance of 4 AU
around a G2V star. Table \ref{data.age} gives the corresponding modal particle sizes in microns (Cooper et al. 2003)
for the ice particles of the water clouds that appear in this evolutionary sequence, as well
as the inner flux \teff, gravity, and planet radius as this 1-\mj EGP evolves according to the theory
of Burrows et al. (1997).  The modal particle size is very roughly constant with age.
The radius of the 1-\mj EGP decreases by $\sim$15\% from 0.1 to 5 Gyr.
Not surprisingly, at a distance of 4 AU no inflection in the $T/P$ profile         
is produced.   Note that at this orbital distance and as early as $\sim$50 Myr (not  shown), water clouds form in Jovian-mass objects.
Note also that at 4 AU, even after 5 Gyr ammonia clouds have not yet formed
in the atmosphere of an irradiated 1-\mj EGP.  This is not true for a similar object in isolation (Burrows, Sudarsky, and Lunine 2003).

Figure \ref{contrastd} depicts the planet-to-star flux ratios from 0.5 \mic to 30 \mic for the orbital
distance study associated with the $T/P$ profiles shown in Fig. \ref{auseq.c}.
In the optical, the flux ratios vary between 10$^{-8}$ and 10$^{-10}$.  In the near infrared,
this ratio varies widely from $\sim$10$^{-4}$ to 10$^{-16}$.  However, in the mid-infrared beyond
10 \mic, the flux ratio varies more narrowly from 10$^{-4}$ to 10$^{-7}$.  Hence, it makes a difference
in what wavelength region one conducts a search for direct planetary light.

Gaseous water absorption features (for all orbits) and methane absorption features (for the outer orbits) sculpt the spectra.
The reflected component due to Rayleigh scattering and clouds (when present) is most manifest
in the optical and the emission component (similar to the spectrum of an isolated low-gravity
brown dwarf) takes over at longer wavelengths.   For the EGPs interior to $\sim$1.0 AU,
the fluxes longward of $\sim$0.8 \mic are primarily due to
thermal emission, not reflection.  These atmospheres do not contain condensates and are
heated efficiently by stellar irradiation.  As a result, the $Z$ ($\sim$1.0 \mic), 
$J$ ($\sim$1.2 \mic), $H$ ($\sim$1.6 \mic), and $K$ ($\sim$2.2 \mic) band
fluxes are larger by up to several orders of magnitude
than those of the more distant EGPs.  
Generally, clouds increase a planet's flux
in the optical, while decreasing it in the $J$, $K$, $L^{\prime}$ ($\sim$3.5 \mic), 
and $M$ ($\sim$5.0 \mic) bands.  The transition between the reflection and emission
components moves to longer wavelengths with increasing distance, and is around 0.8-1.0 \mic at 0.2 AU and
$\sim$3.0 \mic at 15 AU.  However, since the irradiation and atmospheric structure and spectra
are being calculated self-consistently, emission and reflection components are in fact inextricibly intertwined and it is not
conceptually correct to separate them.  

At large distances exterior to $\sim$3.5 AU,
the flux longward of $\sim$15 \mic manifests undulations due to pressure-induced absorption by H$_2$.
Importantly, there is always a significant bump around the $M$ band at 4-5 \mic.  The peak of this bump
shifts from $\sim$4 \mic to $\sim$5 \mic with increasing orbital distance.  Though muted by the presence
of clouds, it is always a prominent feature of irradiated EGPs, as it is in T dwarfs and the
Jovian planets of our solar system.  Curiously, but not unexpectedly, as Fig. \ref{contrastd} indicates, the planet/star flux ratio is
most favorable in the mid-infrared.  This fact should be of some interest to those planning 
TPF or successor missions to the Spitzer Space Telescope.

Though the major trend is a monotonic decrease in a planet's flux with increasing orbital distance,
the 0.2-AU model atmosphere is hot enough that the sodium and potassium
resonance absorption lines appear and surpress the flux around Na-D (0.589 \mic) and 
the related K I doublet at 0.77 \mic.  The result is a lower integrated visible           
flux that is comparable to that of the otherwise dimmer 0.5-AU model.
Figures \ref{contrastd.2} and \ref{contrastd.2b} focus in on the 0.5 \mic to 
2.0 \mic region and allow one to distinguish one model 
from another at shorter wavelengths more easily than is
possible in the panoramic Fig. \ref{contrastd}. 
These figures allow us to see that for greater orbital distances, 
the atmospheric temperatures are too low for the alkali metals to
appear, but the methane features near 0.62 \mic, 0.74 \mic, 0.81 \mic, and 0.89 \mic come into their own.
Broad water bands around 0.94 \mic, 1.15 \mic, 1.5 \mic and 1.85 \mic that help to define the
$Z$, $J$, and $H$ bands are always in evidence, particularly for the 0.2 and 1.0 AU models
that don't contain water clouds.  For greater distances, the presence of water clouds slightly
mutes the variation with wavelength in the planetary spectra. 
Hence, smoothed water features and methane bands
predominate beyond $\sim$1.5 AU.

\section{Age Dependence from 100 Myr to 5 Gyr of the Spectrum of a 1-\mj EGP at a Given Distance}
\label{age_sequence}

\begin{figure} 
\plotone{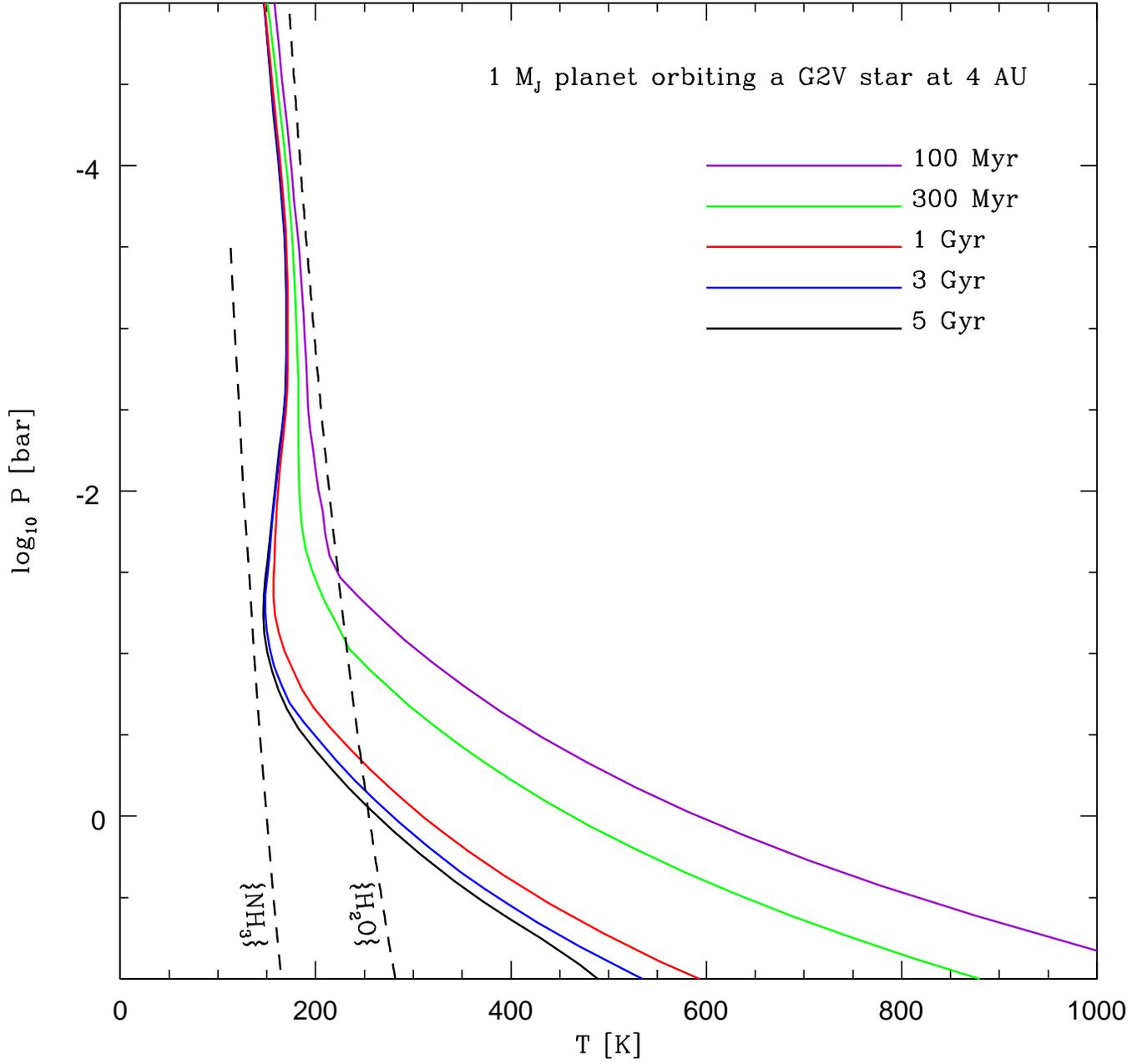}
\caption{Similar to Fig. \ref{auseq.c}, this figure depicts atmospheric profiles 
of temperature (in Kelvin) versus the logarithm base ten of the pressure (in bars)
for a family of irradiated 1-\mj EGPs around a G2V star, but as a function of age at a given orbital distance of 4.0 AU.
The ages vary from 0.1 to 5.0 Gyr.  
Note that the pressure is decreasing along the ordinate, which thereby resembles altitude.
The cloud condensation curves for both NH$_3$ (moot for this sequence) and H$_2$O are given as the dashed lines
and the spectral/atmospheric models include the effects of the water clouds in a consistent way. 
See text in \S\ref{distance_sequence} and \S\ref{age_sequence} for relevant details and discussion.}
\label{ageseq.c}
\end{figure}

Figure \ref{contrasta} presents the planet-to-star flux ratios from 0.5 \mic to 6.0 \mic for a 1-\mj EGP orbiting a G2V star at 4 AU
as a function of age. Figure \ref{ageseq.c} depicts the corresponding $T/P$ profiles, along with the NH$_3$ and H$_2$O
condensation lines at solar metallicity.  The theory of Burrows et al. (1997) is used to obtain an approximate mapping
between mass, age, internal \teff, and gravity\footnote{We note that the star evolves as well, but 
for clarity we have neglected this effect.}.   The depicted ages are 0.1, 0.3, 1, 3, and 5 Gyr. 
These planet parameters are chosen merely to represent the systematics with age;  different EGP masses and
orbital distances will yield quantitatively different spectra.
Table \ref{data.age} shows that for this suite of models
the internal flux \teff varies from 290 K to 103 K, the surface 
gravity varies from 1695 cm s$^{-2}$ to 2325  cm s$^{-2}$,
and the planet radius varies from 1.17 \rj to 1.0 \rj.

As is clear from Fig. \ref{contrasta}, younger EGPs with higher inner boundary \teff{s} (Table 
\ref{data.age}) have much higher fluxes in the  $Z$, $J$, $H$, $K$, and $M$ bands.
However, the older objects, having cooled more, have lower internal luminosities.
This results in lower fluxes in those same bands by as much as two orders of magnitude.  
As Figure \ref{ageseq.c} shows, the older EGPs have progressively
deeper water clouds.  What is not obvious from Figure \ref{ageseq.c} is that these clouds are also thicker.  This results in a very slightly
increasing reflected optical flux with increasing age that accompanies the reverse trend in the near infrared.
Hence, the fluxes in the optical shortward of $\sim$1.0 \mic are only weak functions of age, while the fluxes
in the  $Z$, $J$, $H$, $K$, and $M$ bands at early ages are strong functions of age.  At later ages, the formation of water
clouds moderates the age dependence of the $Z$, $J$, $H$, and $K$ band fluxes.  In fact, there can be 
slight increases in flux in the $Z$, $J$, and $H$ bands with increasing age.   However, the $M$ band
flux continues to be diagnostic of age, monotonically decreasing by almost two orders of magnitude from 0.1
to 5 Gyr.  Hence, the best diagnostics of age are in the near infrared, not the optical.

\section{Irradiated EGP Spectra as a Function of Mass from 0.5\mj to 8\mj at a Given Age and Orbital Distance}
\label{mass_sequence}

Figure \ref{contrastm} portrays planet-to-star flux ratios from 0.4 \mic to 6.0 \mic for a 5-Gyr EGP orbiting a G2V star at 4 AU,
as a function of EGP mass.  The masses represented are 0.5, 1, 2, 4, 6, and 8 \mj.  An inner flux boundary condition
\teff\ from the evolutionary calculations of Burrows et al. (1997) has been employed and is given in Table \ref{data.mass}.
\teff\ varies from 82 K to 251 K, the surface gravity varies from 1290 cm s$^{-2}$ to 17800 cm s$^{-2}$, and the radii
vary from 0.95 \rj to 1.04 \rj.  Note that these radii peak in the middle of the sequence near 4 \mj. 
Due to the 5-Gyr age assumed, all these models have water clouds and the derived modal particle sizes decrease monotonically 
with increasing mass from 146 \mic at 0.5 \mj to 39 \mic at 8 \mj.  In general, the larger
the EGP mass, the higher in the atmosphere the clouds form, but
cloud position is not fully monotonic at the low-mass end of the
sequence.

For higher mass, at a given age an EGP's inner \teff\ and internal luminosity are higher.  
This results in higher fluxes in the $Z$, $J$, $H$, $K$, and $M$
bands for higher masses and is similar to the trend seen in \S\ref{age_sequence} with decreasing age.
However, larger-mass EGPs also have higher surface gravities, which result in water clouds 
with lower column depths, and, hence, lower optical depths, despite the 
contrary trend of modal particle size (Table \ref{data.mass}).
The upshot is that higher-mass EGPs have slightly smaller planet/star flux ratios
in the optical.  This optical component is due to reflection off of cloudy atmospheres 
with roughly similar compositions.  Hence, there is an anti-correlation
between flux levels in the visible and near-infrared that might be 
diagnostic of planet mass for Jupiter-aged EGPs.

\section{Irradiated Brown Dwarfs}
\label{browndwarfs}

\begin{figure}
\plotone{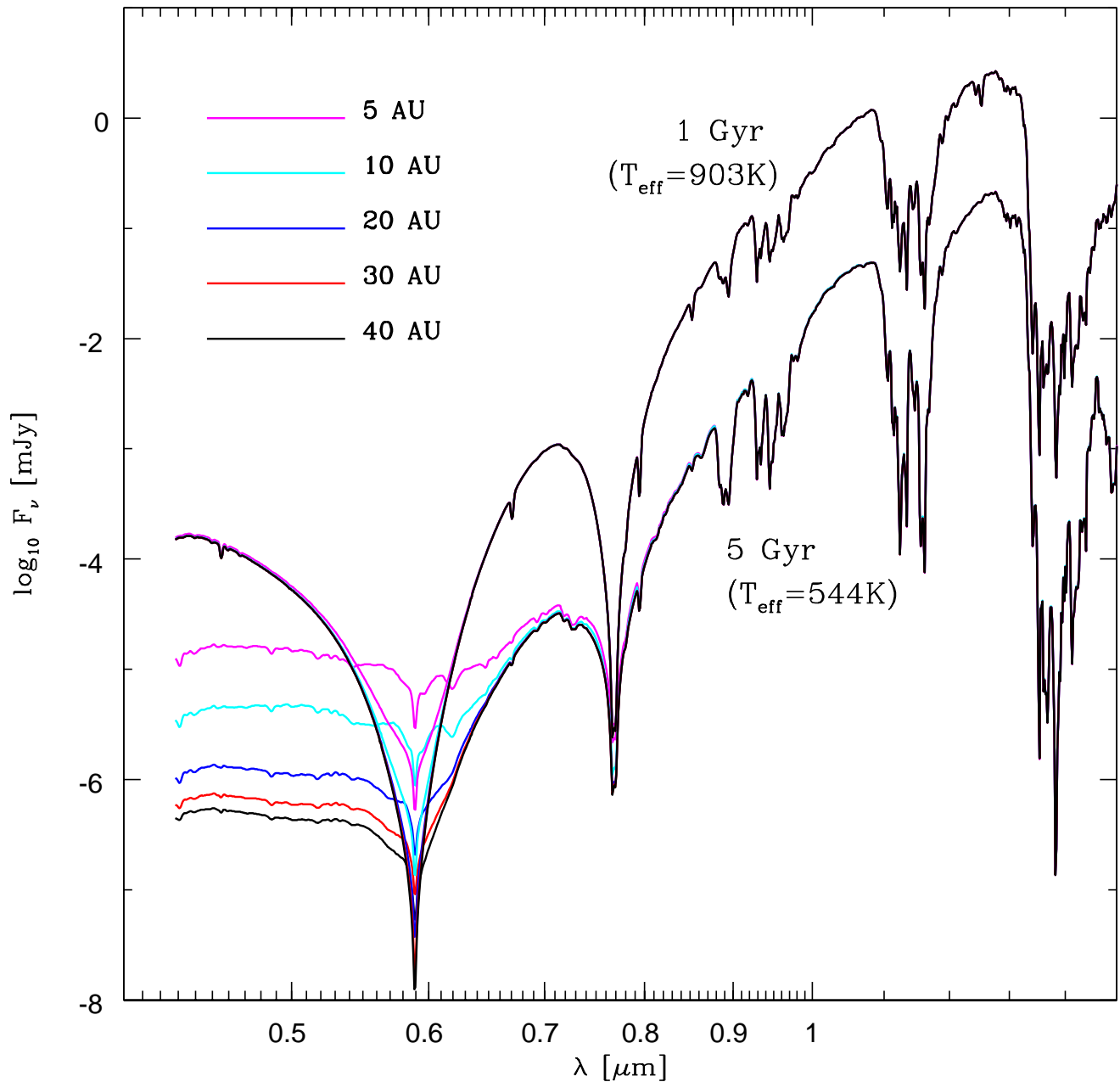}
\caption{Depicted are theoretical spectra of a 30-\mj brown dwarf at ages of 1 and 5 Gyr in
orbit around a G2V star that is irradiating it.  The logarithm base ten of the flux in milliJanskys at 10 parsecs
versus wavelength in microns from 0.4 \mic to 1.5 \mic is given.  The theory of Burrows et al. (1997) was
used to determine \teff\ and gravity for these ages and mass.  The results are 
shown for different orbital distances from 5 to 40 AU.
See text for a discussion.} 
\label{brown}
\end{figure}

Depicted in Fig. \ref{brown} are theoretical spectra of a 30-\mj brown dwarf at ages of 1 and 5 Gyr in
orbit around a G2V star from 0.4 \mic to 1.5 \mic.  The results are for orbital distances from 5 AU to 40 AU
and a distance to the Earth of 10 parsecs and include irradiation effects.  As before, the theory of Burrows et al. (1997) is
used to determine \teff\ and gravity for this mass and these ages. 
In Fig. \ref{brown}, the prominence of the Na-D (0.589 \mic) and K I (0.77 \mic) features at shorter 
wavelengths is clear and is canonical for brown dwarfs (Burrows, Marley, and Sharp 2000).  

Unlike for the lower mass EGPs discussed in \S\ref{distance_sequence}, \S\ref{age_sequence}, and \S\ref{mass_sequence},
the internal luminosity of such a relatively massive SMO dominates its energy budget.  As a consequence, the brown dwarf's
spectrum longward of 0.9 \mic is unaffected by irradiation.  However, the reflected component in the optical, particularly
for the older brown dwarf with lower internal heat content and luminosity, is a function of distance.  As Fig. \ref{brown} demonstrates,
the optical flux from a cool brown dwarf can be elevated shortward of 0.7 \mic by as much as an order
of magnitude in the $V$ band.  In particular, the shape of the Na-D feature can be significantly altered.
Since there are no clouds in the atmospheres of these brown dwarf models,
Rayleigh scattering off of H$_2$, He, and H$_2$O accounts for this reflection.
Our prediction is that the optical spectra of brown dwarfs that are close companions to K, G, or F stars
will be modified by irradiation.  Note that the T dwarf Gliese 229B is at a projected distance of $\sim$40 AU,
but that its primary is an M4V star.  Such a star is too dim to so radically alter a brown dwarf's optical flux.
For an L dwarf companion, the presence of silicate clouds in its atmosphere will reflect a primary's light 
in the blue and UV.  Someday, such a reflected component might be detectable.

\section{Predicted Phase-Averaged Spectra for Known EGPs at Wide Angular Separations}
\label{specific_EGPs}

Table \ref{data.angular} lists many of the known EGPs that, due to a propitious combination
of semi-major axis and distance from the Earth, are at wide angular separations from their 
parent stars\footnote{For comparison, some of the closest EGPs are also shown at the bottom of Table \ref{data.angular}.}.  
Ordered by decreasing separation (defined as the ratio of semi-major axis to distance), 
Table \ref{data.angular} also lists the stellar type of the primary, semi-major axis, Hipparcos distance,
orbital period, $m_p\sin(i)$, and orbital eccentricity.  This family of known EGP comprises some
of the prime candidates for direct detection of planetary light using interferometric, 
adaptive-optics, or coronagraphic techniques (\S\ref{intro}).
Note that the separations quoted in Table \ref{data.angular} don't take into account  
the projection of the orbit or variations due to non-zero eccentricities (which can be large).
In particular, variations due to the significant excursions in planet-star distance
that attend large eccentricities can result in large changes in irradiation regimes.
In turn, this can result in large variations in planet spectrum.  Interestingly, it
is possible for an EGP atmosphere to cycle between states with and without clouds,
with the concomitant large changes in flux ratios and spectral signatures with orbital phase 
(Sudarsky, Burrows, and Hubeny 2004).  Figure \ref{contrastd} gives some idea
of the range of spectral variation possible for highly-eccentric EGPs.

Figures \ref{contrastd}, \ref{contrasta}, and \ref{contrastm} provide a broad-brush view
of generic EGP spectra for a range of orbital distances, ages, and masses.  With Figs. \ref{c39}, \ref{c77},
and \ref{c47}, we provide phase-averaged predictions/calculations for a specific subset 
of the known EGPs listed in Table \ref{data.angular}.  This subset includes HD 39091b, 
$\gamma$ Cephei b, HD 70642b, $\upsilon$ And d, Gliese 777A b, HD 216437b, HD 147513b, 55 Cancri d, 47 UMa b, 47 UMa c, 
14 Her b, and $\epsilon$ Eri b.  Table \ref{data.specific} itemizes these EGPs,
along with their derived \teff{s} and surface gravities.  In addition, Table \ref{data.specific} 
lists the theoretical modal particle sizes of the water droplets
that form in their atmospheres.   For definiteness, the planetary masses are set 
equal to the measured $m_p\sin(i)$ and, as before, the evolutionary theory of Burrows et al. (1997)
is used to estimate the corresponding surface gravities and inner boundary \teff{s}. 
As described in SBH and \S\ref{numerical}, a stellar spectrum
from Kurucz (1994) for the primary stellar type given for each EGP in Table \ref{data.specific} is used  
for the outer irradiation boundary condition in the self-consistent atmosphere/spectrum calculation.
As a comparison of Figs. \ref{contrastd}, \ref{contrastd.2}, and \ref{contrastd.2b} 
with Figs. \ref{c39}, \ref{c77}, and \ref{c47} demonstrates, 
the major dependence of the planet-to-star flux ratio is with orbital distance.
Note that for all of the EGPs listed in Table \ref{data.specific}, the assumption that the 
planet-star distance is fixed at the measured semi-major axis 
does not result in the formation of ammonia clouds.  However, 
such clouds should form near apastron for the known EGPs 
with large semi-major axes and high eccentricities, such as 
$\epsilon$ Eri b and 55 Cnc d.  The appearance and disappearance of such clouds
with orbital phase would be exciting signatures to detect.


\section{Summary and Conclusions}
\label{conclusion}

In this paper, we have calculated theoretical phase-averaged planet/star flux ratios
in the optical, near-infrared, and mid-infrared for wide-separation irradiated EGPs as
a function of orbital distance, mass, and age.  We have also predicted the 
corresponding quantities for 12 specific known EGPs, given their corresponding
primary spectra, average orbital distances, approximate masses, and approximate ages.
Hence, we have explored various physical diagnostics that can inform the
direct EGP detection programs now being planned
or proposed.  In the optical, the flux ratios for distances from 0.2 AU to 15 AU 
and masses from 0.5 \mj to 8 \mj can vary from slightly above 10$^{-8}$ to $\sim$10$^{-10}$.
In the near infrared around 2-4 \mic, the flux ratio ranges more widely, spanning values
from 10$^{-4}$ to as low as 10$^{-16}$.  At $\sim$ 5 \mic, the planet/star flux
ratio ranges approximately four orders of magnitude and can be as high as 10$^{-4}$.
The $M$ band should be a useful region to explore and $M$-band fluxes
are sensitive to the presence of clouds.  For all models, the mid-infrared from 10 \mic to
30 \mic is always encouragingly high.  In fact, for closer separations (0.05 AU-0.1 AU), not
the subject of this paper, the flux ratio at $\sim$20 \mic can be $\sim$$10^{-3}$. 

Depending upon orbital distance, age, and mass, spectral features due to methane, water, and the alkali
metals are prominent.  Furthermore, there is a slight anti-correlation in the effects
of clouds in the optical and infrared, with the optical fluxes increasing 
and the infrared fluxes decreasing with increasing cloud depth.  
For young and massive irradiated EGPs, there are prominent peaks in the $Z$, $J$, $H$, $K$, and $M$ bands.
Though the optical flux is not a very sensitive function of age, there is a useful age dependence
of the fluxes in these bands.  Furthermore, there is an anti-correlation 
with increasing mass at a given age and orbital distance
between the change in flux in the optical and in the near-IR bands, with the optical
fluxes decreasing and the $Z$, $J$, $H$, $K$, and $M$ band fluxes increasing with increasing mass.

The spectra of more massive SMOs (brown dwarfs) longward of $\sim$0.9 \mic are not 
significantly affected by stellar irradiation.  Their internal heat content and interior fluxes
are too large.  However, brown dwarf fluxes from 0.4 \mic to 0.65 \mic, 
can be enhanced by Rayleigh reflection by as much as a factor of 10.  This is particularly true of old
or low-mass brown dwarfs and is a predictable function of distance.  Moreover, irradiation 
can alter the profile shape of the Na-D feature significantly.  

We are not able to calculate EGP spectra for all possible combinations
of mass, age, composition, orbital distance, eccentricity, orbital phase,
Keplerian element, and primary spectral type.  This fact is what motivates the more 
modest synoptic view we have provided in this paper.  However, in the process of 
developing the tools for this study, we have established the capability to calculate
EGP spectra for any combination of these parameters. In particular, 
Sudarsky, Burrows, and Hubeny (2004) address the orbital phase and eccentricity
dependences of irradiated EGPs spectra.

\begin{figure}
\plotone{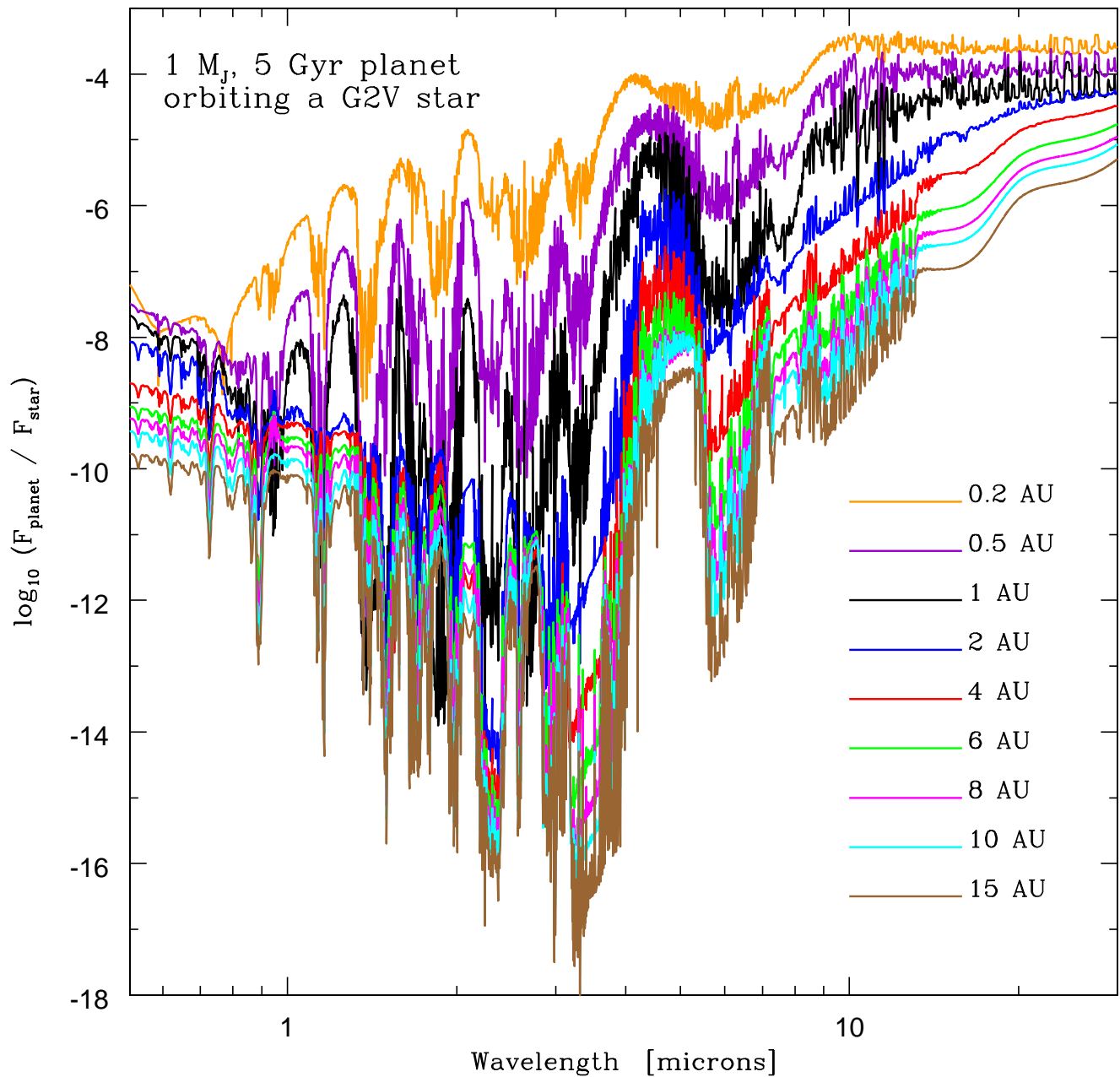}
\caption{Planet to star flux ratios versus wavelength (in microns) from 0.5 \mic to 30 \mic
for a 1-\mj EGP with an age of 5 Gyr orbiting a G2V main sequence star similar to the Sun.
This figure portrays ratio spectra as a function of orbital diatance from 0.2 AU
to 15 AU.  Zero eccentricity is assumed and the planet spectra have been phase-averaged
as described in Sudarsky, Burrows, and Hubeny (2003). 
The associated $T/P$ profiles are given in Fig. \ref{auseq.c} and Table \ref{data.dist}
lists the modal radii for the particles in the water and ammonia clouds.
Note that the planet/star flux ratio is most favorable in the 
mid-infrared.  See text for further discussion.}
\label{contrastd}
\end{figure}

\begin{figure}
\plotone{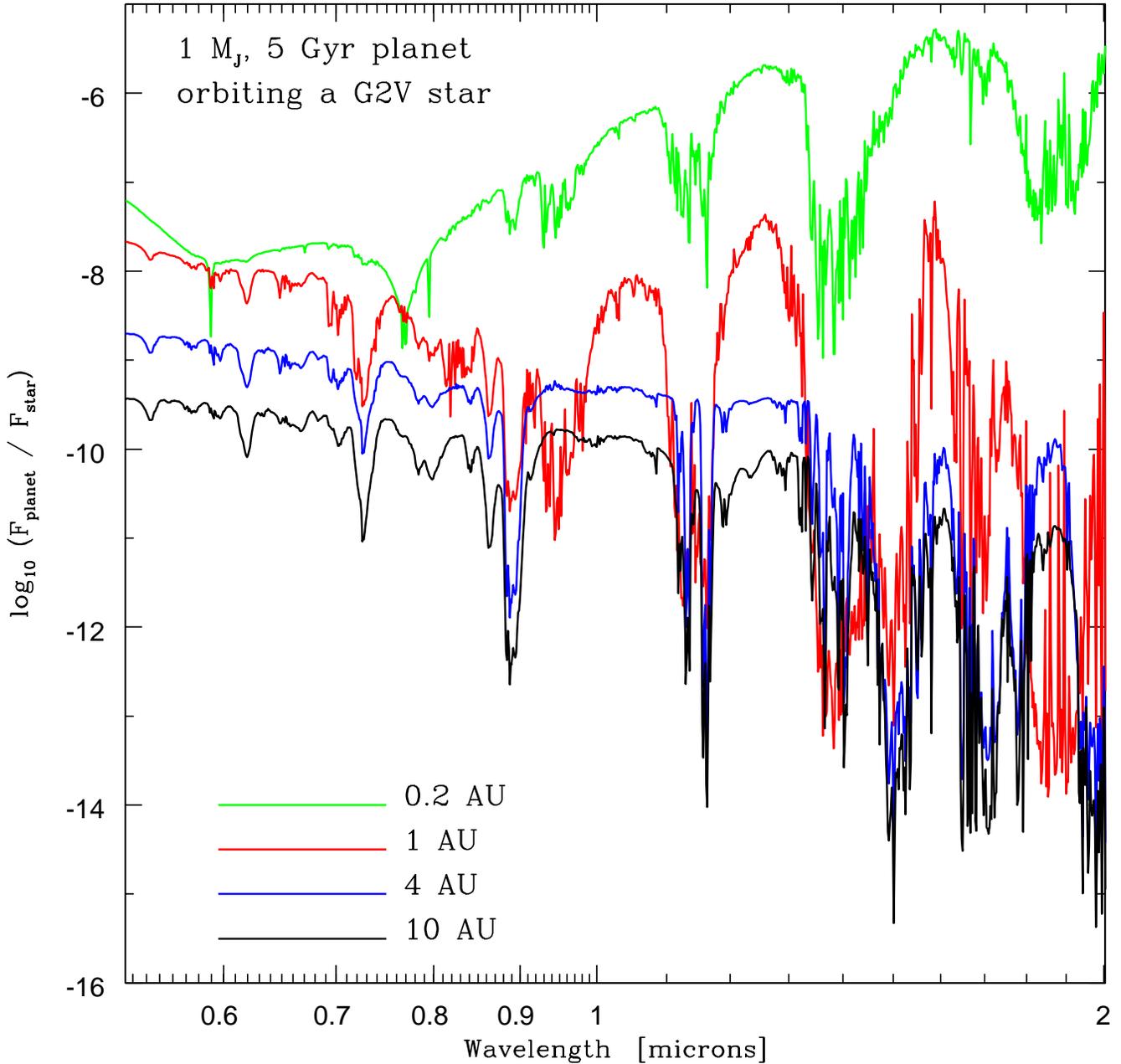}
\caption{The same as Fig. \ref{contrastd}, but highlighting the shorter wavelengths 
and for a subset of distances (0.2, 1, 4, 10 AU).   This figure provides a clearer
picture of the features shortward of 2.0 \mic for each of the represented models.  For the 0.2 AU model, 
the temperatures of the atmosphere are high enough for the Na-D doublet around 0.589 \mic and the K I 
doublet near 0.77 \mic to be visible.  These features are even more prominent for closer-in EGPs (Sudarsky, Burrows, and Hubeny 2003).
At greater orbital distances, the atmospheric temperatures are too low for the alkali metals to
appear, but the methane features near 0.62 \mic, 0.74 \mic, 0.81 \mic, and 0.89 \mic come into their own.
Water bands around 0.94 \mic, 1.15 \mic, 1.5 \mic and 1.85 \mic that help to define the
$Z$, $J$, and $H$ bands are always of importance. 
For greater distances, the presence of water clouds 
can smooth the variations in the planetary spectra that would otherwise be large due to the
strong absorption features of gaseous water vapor.
See text for discussion.}
\label{contrastd.2}
\end{figure}

\begin{figure}
\plotone{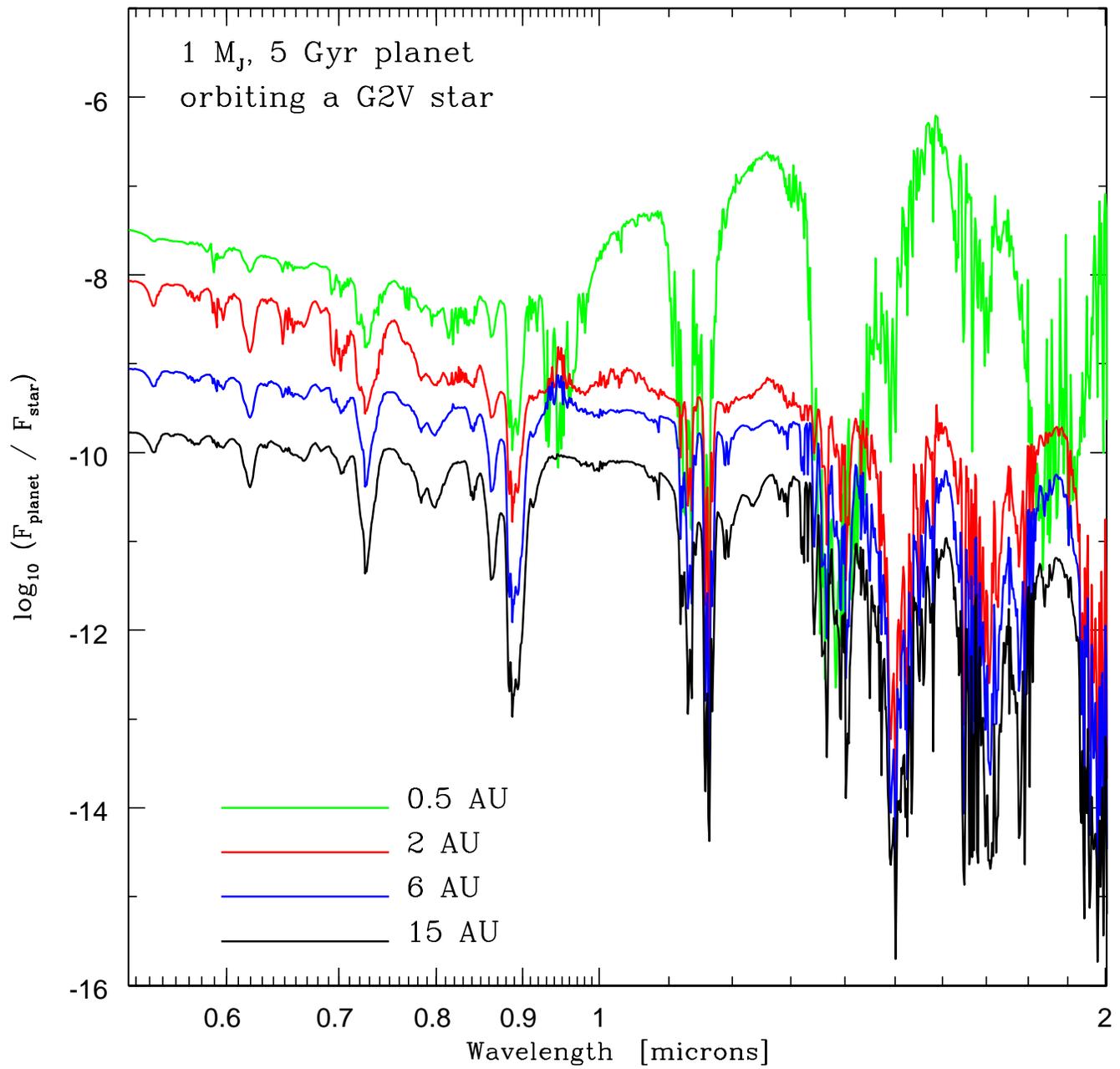}
\caption{The same as Fig. \ref{contrastd}, but, as in Fig. \ref{contrastd.2}, highlighting the shorter wavelengths.
A different subset of distances (0.5, 2, 6, 15 AU) is shown.  See the text and the figure caption for Fig. \ref{contrastd.2}
for details.  Figures \ref{contrastd.2} and \ref{contrastd.2b} allow one to distinguish more easily than is
possible in the panoramic Fig. \ref{contrastd} one model from another. }
\label{contrastd.2b}
\end{figure}

\begin{figure}
\plotone{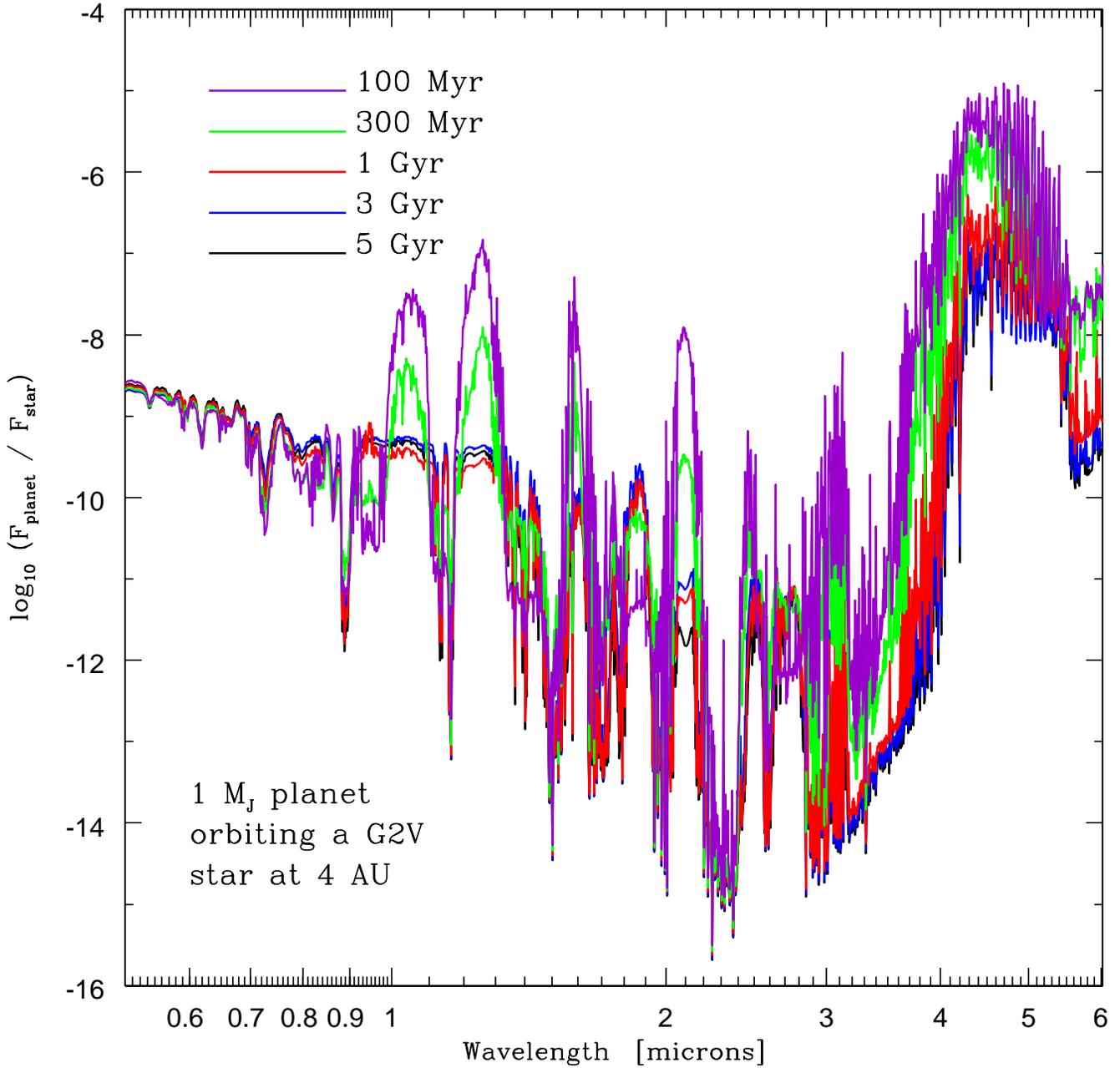}
\caption{The planet-to-star flux ratio from 0.5 \mic to 6.0 \mic for a 1-\mj EGP orbiting a G2V star at 4 AU
as a function of age.  The ages are 0.1, 0.3, 1, 3, and 5 Gyr.  An inner flux boundary condition
\teff from the evolutionary calculations of Burrows et al. (1997) has been employed.
The effect of clouds is handled in the radiative transfer calculation in a completely consistent fashion.
See Table \ref{data.age}, Figure \ref{ageseq.c}, and text for details and discussion.}
\label{contrasta}
\end{figure}

\begin{figure}
\plotone{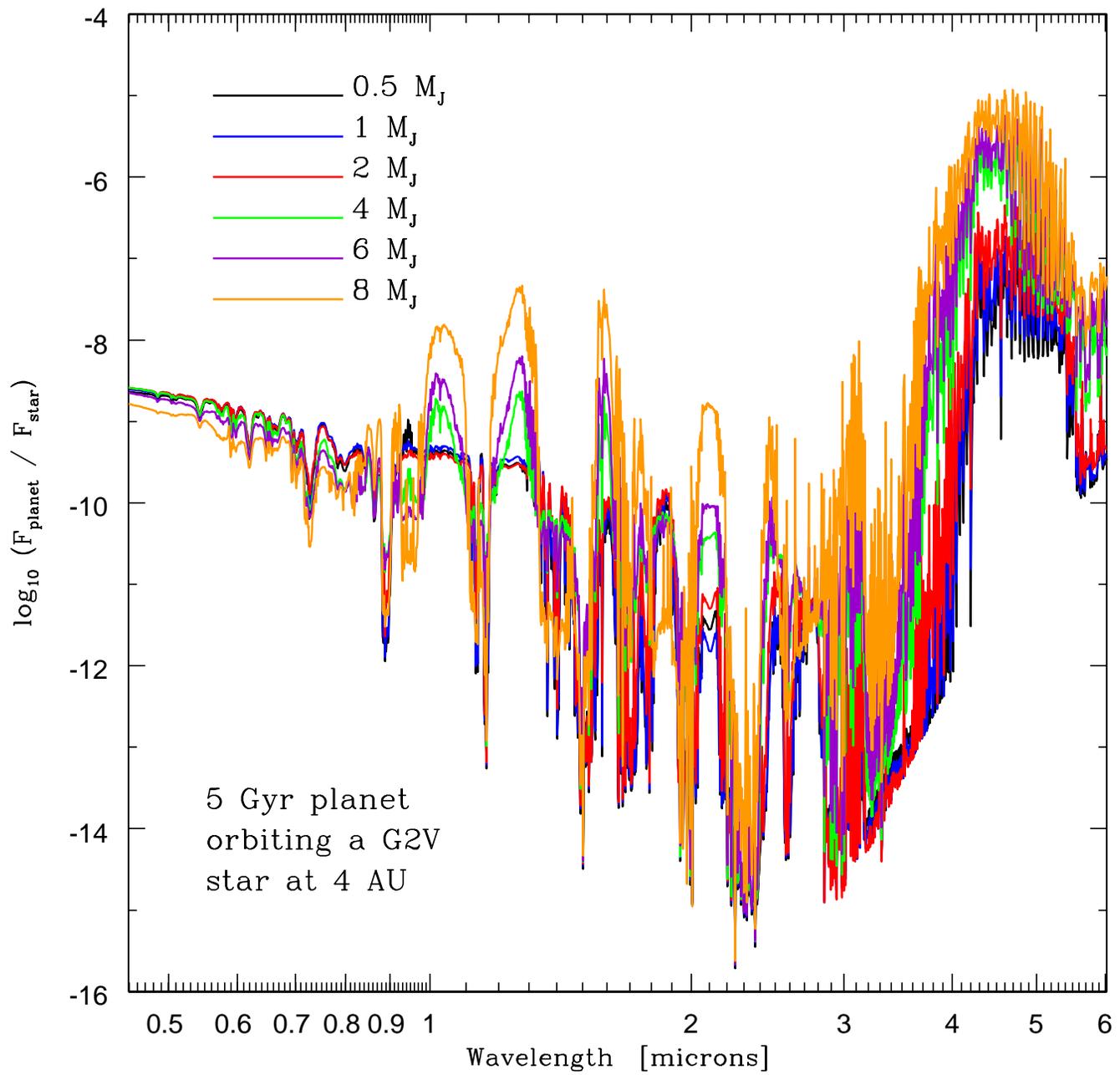}
\caption{Similar to Fig. \ref{contrasta}, but the planet-to-star flux ratio from 0.4 \mic to 6.0 \mic for a 5-Gyr EGP orbiting a G2V star at 4 AU,
as a function of EGP mass.  The masses represented are 0.5, 1, 2, 4, 6, and 8 \mj. 
See Table \ref{data.mass} and text for a discussion.}
\label{contrastm}
\end{figure}

\begin{figure}
\plotone{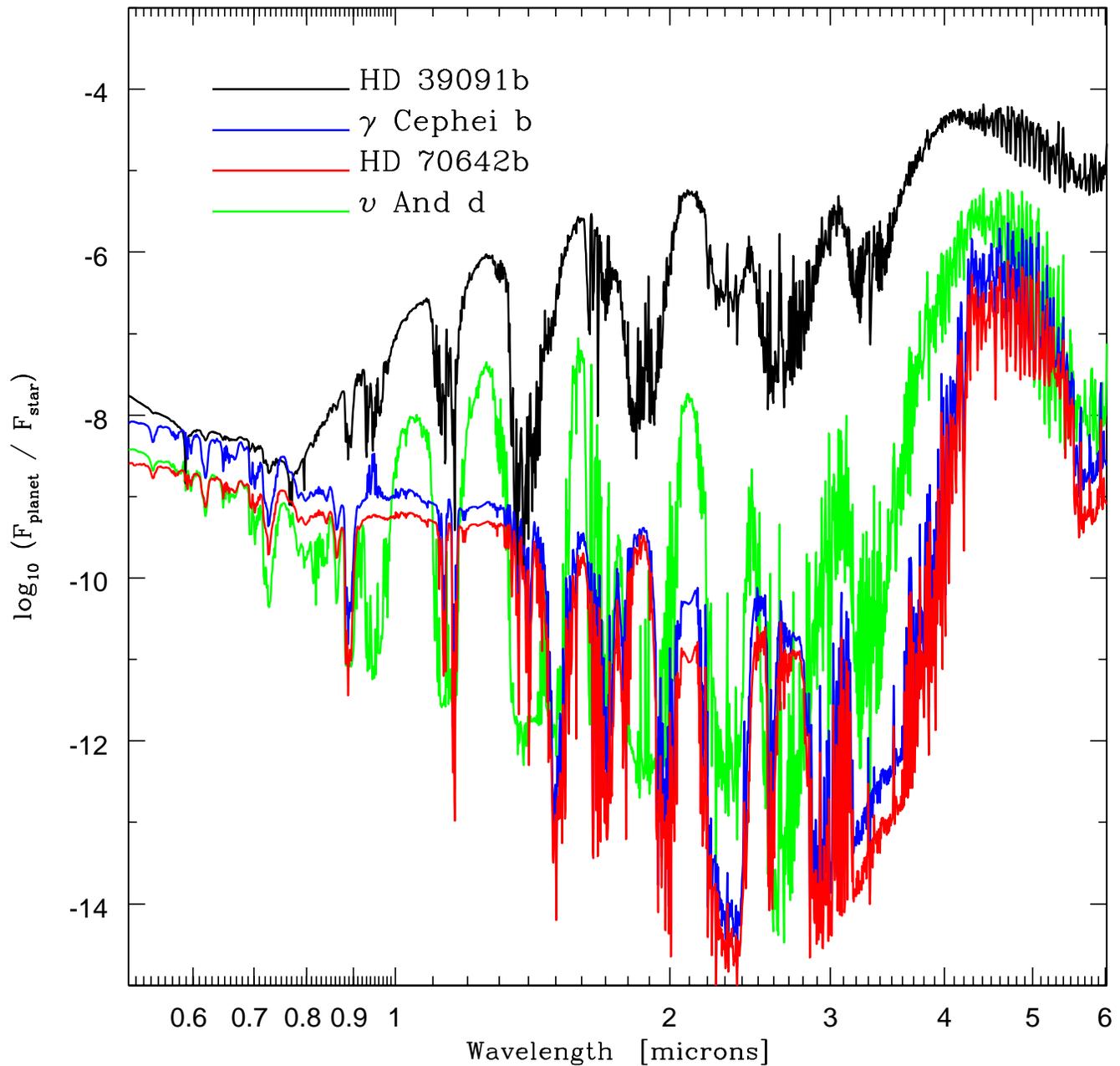}
\caption{Similar to Figs. \ref{contrasta} and \ref{contrastm}, but for four specific known EGPs listed in Table \ref{data.specific}.  
These are HD 39091b, $\gamma$ Cephei b, HD 70642b, and $\upsilon$ And d.
The wavelengths range from 0.5 \mic to 6.0 \mic.
The planet fluxes are phase-averaged and the effect of the known eccentricities
is ignored.  The orbital distances are assumed to be equal to the measured 
semi-major axes and the planets' masses are set equal to the measured values of $m_p\sin(i)$. For all of these objects
water clouds are formed and the modal particle sizes are given in Table \ref{data.specific}.   Table \ref{data.specific}
also shows the values of the gravity and the \teff\ that describes the inner flux, both derived using the theory of Burrows et al. (1997).  
Compare this figure with Figs. \ref{contrastd}, \ref{contrasta}, and \ref{contrastm} and see text for details.}
\label{c39}
\end{figure}

\begin{figure}
\plotone{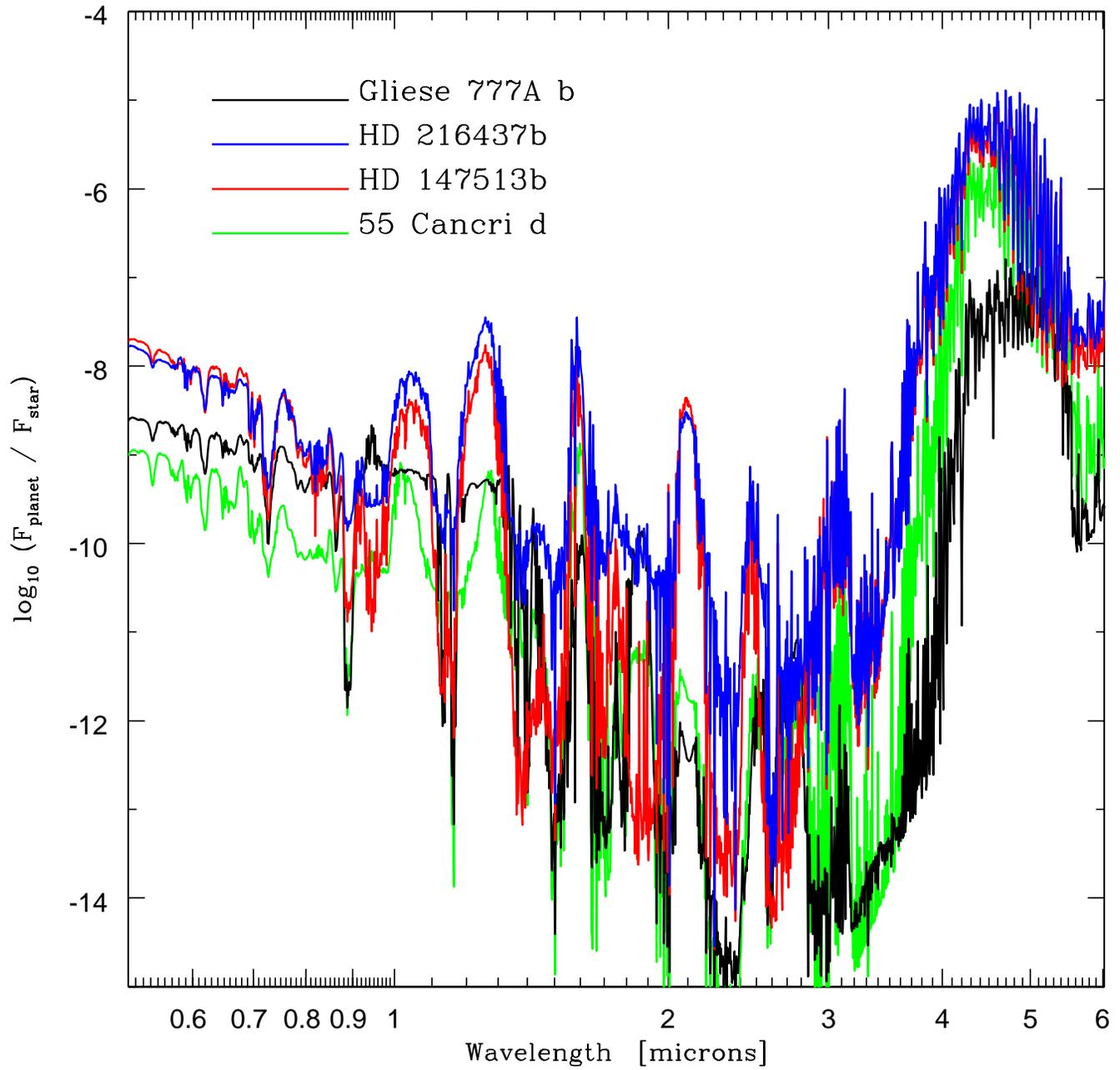}
\caption{Same as Fig. \ref{c39}, but for Gliese 777A b, HD 216437b, HD 147513b, and 55 Cancri d.
Refer to Fig. \ref{c39}, Table \ref{data.specific}, and the text for further details and discussion.}
\label{c77}
\end{figure}

\begin{figure}
\plotone{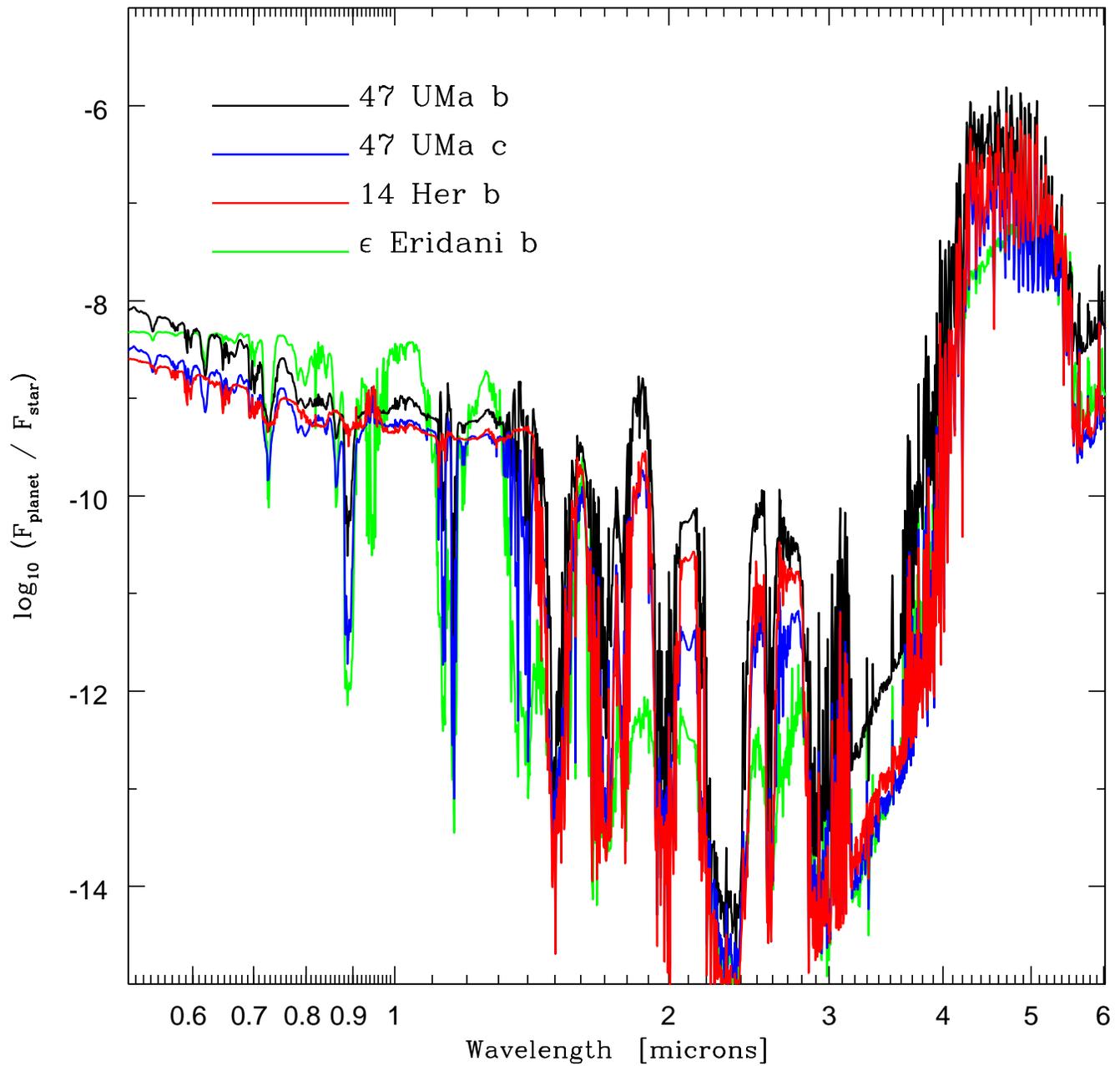}
\caption{Same as Figs. \ref{c39} and \ref{c77}, but for 47 UMa b, 47 UMa c, 14 Her b, and $\epsilon$ Eridani b. 
Refer to Fig. \ref{c39}, Table \ref{data.specific}, and the text for further details and discussion.} 
\label{c47}
\end{figure}

The remote sensing of the atmospheres of EGPs will be challenging, but the 
detection and characterization of the direct light from a planet outside
our solar system will be an important milestone in both astronomy and planetary 
science.  One instrument proposed to meet this challenge is the space-based coronagraphic imager
{\it Eclipse} (Trauger et al. 2000,2001).  The {\it Eclipse} instrument team is predicting a contrast
capability of $\sim$10$^{-9}$ for an inner working angle of
0.3$^{\prime\prime}$ or 0.46$^{\prime\prime}$ in the $V$/$R$ or $Z$ bands, respectively.   
As Figs. \ref{contrastd} through \ref{contrastm} indicate, with such a capability
irradiated EGPs could be detected and analyzed.  Many could even be discovered.
However, whether such sensitivity is achievable remains to
be demonstrated.  Be that as it may, further advancement in our understanding of
extrasolar planets is contingent upon technical advances that would enable the direct measurement
of the dynamically dominant and brighter components of extrasolar planetary systems, the EGPs.

\acknowledgments
 
The authors wish to acknowledge Bill Hubbard, Jonathan Lunine,
Jim Liebert, John Trauger, Jonathan Fortney, Aigen Li, Christopher Sharp, Drew Milsom,
Maxim Volobuyev, and Curtis Cooper
for fruitful conversations or technical aid and help during the
course of this work, as well as
NASA for its financial support via grants NAG5-10760
and NAG5-10629.  Furthermore, we acknowledge support through the Cooperative Agreement \#{NNA04CC07A} 
between the University of Arizona/NOAO LAPLACE node and NASA's
Astrobiology Institute.  
Finally, the first author would also like to 
thank the Kavli Institute for Theoretical Physics where 
some of this work was performed.

{}

\begin{deluxetable}{ccc}
\tablewidth{7cm}
\tablecaption{Model Data for Distance Sequence\label{data.dist}}
\tablehead{
\colhead{$a$ (AU)}  & \colhead{condensate(s)}  &  \colhead{$r_0$ ($\mu$m)}}
\startdata

  $0.2$& none & - \\
  $0.5$& none & - \\
  $1$& none & - \\
  $2$& H$_2$O & 107 \\
  $4$& H$_2$O & 109 \\
  $6$& NH$_3$, H$_2$O & 56, 110 \\
  $8$& NH$_3$, H$_2$O & 56, 111 \\
 $10$& NH$_3$, H$_2$O & 57, 111 \\
 $15$& NH$_3$, H$_2$O & 62, 111 \\

\enddata
\end{deluxetable}

\begin{deluxetable}{ccccc}
\tablewidth{11cm}
\tablecaption{Model Data for Age Sequence\label{data.age}}
\tablehead{
\colhead{Age (Gyr)}  & \colhead{T$_{\textrm{eff}}$ (K)} &  \colhead{$log_{10}$ $g$ (cm s$^{-2}$)} & \colhead{$R/R_J$}  & \colhead{$r_0$ ($\mu$m)}}
\startdata

  $0.1$& 290 & 3.23 & 1.17 & 123 \\
  $0.3$& 219 & 3.27 & 1.11 & 119 \\
  $1$&  159 & 3.31 & 1.06 & 123 \\
  $3$&  118 & 3.35 & 1.02 & 111 \\
  $5$&  103 & 3.37 & 1.00 & 109 \\

\enddata
\end{deluxetable}

\begin{deluxetable}{ccccc}
\tablewidth{10.5cm}
\tablecaption{Model Data for Mass Sequence\label{data.mass}}
\tablehead{
\colhead{$M/M_J$}  & \colhead{T$_{\textrm{eff}}$ (K)} &  \colhead{$log_{10}$ $g$ (cm s$^{-2}$)} & \colhead{$R/R_J$}  & \colhead{$r_0$ ($\mu$m)}}
\startdata

  $0.5$& 82 & 3.11 & 0.95 & 146 \\
  $1$&  103 & 3.37 & 1.00 & 109 \\
  $2$&  134 & 3.64 & 1.03 & 80 \\
  $4$&  177 & 3.93 & 1.04 & 57 \\
  $6$&  216 & 4.12 & 1.03 & 46 \\
  $8$&  251 & 4.25 & 1.02 & 39 \\

\enddata
\end{deluxetable}

\begin{deluxetable}{cccccccc}
\tablewidth{16.5cm}
\tablecaption{Interesting EGPs Listed by Angular Separation\label{data.angular}}
\tablehead{
\colhead{EGP} & \colhead{separation ($^{\prime\prime})$} & \colhead{star}
& \colhead{a (AU)} & \colhead{d (pc)}
& \colhead{P} & \colhead{Msin$i$ (\mj)} & \colhead{$e$}}
\startdata

$\epsilon$ Eri b & 1.0 & K2V & 3.3 & 3.2  & 6.85 yrs. & 0.86 & 0.61 \\
55 Cnc d          & 0.44 & G8V & 5.9 & 13.4 & 14.7 & 4.05 & 0.16  \\
47 UMa c          & 0.28 & G0V& 3.73& 13.3 & 7.10 & 0.76 & 0.1 \\
Gl 777A b         & 0.23& G6V& 3.65& 15.9 & 7.15 & 1.15 & $\sim$0 \\
$\upsilon$ And d  & 0.19 & F8V& 2.50& 13.5 & 3.47 & 4.61 & 0.41 \\
HD 39091b         & 0.16& G1IV& 3.34& 20.6 & 5.70 & 10.3 & 0.62 \\
47 UMa b          & 0.16 & G0V& 2.09& 13.3 & 2.98 & 2.54 & 0.06 \\
$\gamma$ Cephei b & 0.15 & K2V& 1.8& 11.8 & 2.5 & 1.25 & $\sim$0 \\
HD 160691c        & 0.15 & G3IV-V&2.3&15.3& 3.56&$\sim$1&$\sim$0.8\\
14 Her b          & 0.15 & K0V& 2.5 & 17   & 4.51 & 3.3  & 0.33 \\
HD 33636b         & 0.12 & G0V& 3.56 & 28.7 & 4.43 & 7.71 & 0.41 \\
HD 10647b         & 0.12 & F9V& 2.10 & 17.3 & 2.89 & 1.17 & 0.32 \\
HD 70642b         & 0.11 & G5IV-V&3.3& 29   & 4.79 & 2.0  & 0.10 \\
HD 216437b        & 0.10 & G4V & 2.7 & 26.5 & 3.54 & 2.1  & 0.34 \\
HD 147513b        & 0.098 & G3V& 1.26& 12.9 & 1.48 & 1.0  & 0.52 \\
HD 160691b        & 0.097& G3IV-V&1.48&15.3 & 1.74 & 1.7  & 0.31 \\
HD 168443c        & 0.087 & G5V& 2.87& 33   & 4.76 & 17.1 & 0.23 \\
HD 50554b         & 0.077 & F8V& 2.38& 31.03& 3.50 & 4.9  & 0.42 \\
HD 106252b        & 0.070 & G0V& 2.61& 37.44& 4.11 & 6.81 & 0.54 \\
HD 10697b         & 0.067& G5IV& 2.0 & 30   & 2.99 & 6.59 & 0.12 \\

\\
$\upsilon$ And c  & 0.061 & F8V & 0.83 & 13.5 & 241 days & 2.11 & 0.18 \\
GJ 876b           &0.045 & M4V & 0.21 & 4.72 & 61.0 & 1.89 & 0.1 \\
GJ 876c           &0.028 & M4V & 0.13 & 4.72 & 30.1 & 0.56 & 0.27 \\
HD 114762b        &0.013& F9V&0.35 & 28 & 84.0 & 11.0 & 0.34 \\
55 Cnc b          &$8.2\times 10^{-3}$&G8V& 0.12 & 13.4 & 14.7 & 0.84 & 0.02 \\
$\upsilon$ And b  &$4.4\times 10^{-3}$&F8V& 0.059 & 13.5 & 4.62 & 0.71 & 0.034 \\
51 Peg b          &$3.4\times 10^{-3}$&G2V& 0.05 & 14.7 & 4.23 & 0.44 & 0.01 \\
$\tau$ Boo b      &$3.3\times 10^{-3}$&F7V& 0.05 & 15 & 3.31 & 4.09 & $\sim$0 \\
HD 209458b        &$9.6\times 10^{-4}$&G0V& 0.045& 47 & 3.52 & 0.69 & $\sim$0 \\
HD 83443b         &$8.7\times 10^{-4}$& K0V&0.038& 43.5 & 2.99 & 0.35 & 0.08 \\

\enddata
\end{deluxetable}

\begin{deluxetable}{cccccccc}
\tablewidth{15.5cm}
\tablecaption{Model Data for Specific EGPs\label{data.specific}}
\tablehead{
\colhead{EGP} &\colhead{star}& \colhead{$a$ (AU)} & \colhead{d (pc)}
& \colhead{$M$sin$i$} & \colhead{T$_{\textrm{eff}}$ (K)} &
\colhead{$log_{10}$ $g$ (cgs)} & \colhead{$r_0$ ($\mu$m)}}
\startdata

$\epsilon$ Eri b & K2V & 3.3 & 3.2 & 0.86 & 170 & 3.48 & 5\tablenotemark{1}\\
55 Cnc d         & G8V & 5.9 & 13.4& 4.05 & 200 & 4.30 & 5\tablenotemark{1}\\
47 UMa c         & G0V & 3.73& 13.3& 0.76 & 95  & 3.48 & 96 \\
Gl 777A b        & G6V & 3.65& 15.9& 1.15 & 100 & 3.48 & 97 \\
$\upsilon$ And d & F8V & 2.50& 13.5& 4.61 & 250 & 4.30 & 5\tablenotemark{1}\\
HD 39091b        & G1IV& 3.34& 20.6& 10.3 & 300 & 4.48 & - \\
47 UMa b         & G0V & 2.09& 13.3& 2.54 & 150 & 3.78 & 67 \\
$\gamma$ Cephei b& K2V & 1.8 & 11.8& 1.25 & 150 & 3.60 & 83 \\
14 Her b         & K0V & 2.5 & 17  & 3.3  & 150 & 3.90 & 59 \\
HD 70642b        & G5V & 3.3 & 29  & 2.0  & 130 & 3.64 & 80 \\
HD 216437b       & G4V & 2.7 & 26.5& 2.1  & 150 & 3.70 & 72 \\
HD 147513b       & G3V & 1.26& 12.9& 1.0  & 100 & 3.37 & 96 \\
\tablerefs{ 1) From Sudarsky, Burrows, \& Hubeny (2003), where a modal
particle size of 5 $\mu$m was assumed.}
\enddata
\end{deluxetable}

\newpage

\newpage

\newpage

\newpage

\newpage

\newpage

\newpage

\newpage

\newpage

\newpage

\newpage

\end{document}